\documentclass[conference]{IEEEtran}
\newcommand{\naive}{Na\"{i}ve }

\ifCLASSINFOpdf
\else
\fi
\usepackage[cmex10]{amsmath}
\usepackage[absolute]{textpos}

\TPMargin{5pt}

\newcommand{\copyrightstatement}{
   \begin{textblock}{0.8}(0.10,0.95)    
        \noindent
        \footnotesize
        Alan Ferrari, Dario Gallucci ``B-Maniacs'', in E. Baccelli, F.
        Juraschek, O. Hahm, T. C. Schmidt, H. Will, M.~W\"ahlisch (Eds.),
        Proc. of 3rd MANIAC Challenge, Berlin, Germany, July 27 - 28, 2013,
        arXiv, Jan. 2014.
        
   \end{textblock}
}

\begin{document}
%
\title{B-Maniac}

\author{\IEEEauthorblockN{Alan Ferrari and Dario Gallucci}
\IEEEauthorblockA{Networking Laboratory, University of Applied Sciences of Southern Switzerland (SUPSI)\\
Manno, Switzerland\\
Email: alan.ferrari@supsi.ch, dario.gallucci@supsi.ch}}


%


\maketitle

\begin{abstract}
Mobile Data Offloading permits users to use cheap communication media, whenever feasible, for delivering their personal data instead of using the infrastructure which is more expensive.  Having good procedures to assess and compare the different possibilities a device has to send his data is crucial. In the following paper, we propose an evaluating approach that takes into consideration the topology changes history in order to provide an efficient way to calculate the quality of a specific medium.\\
In particular \textit{Bayesian Networks} are used as basis to provide our solution to the Maniac Challenge 2013. Bayesian Networks is a statistical model for the generation of inferences starting with very little information. In the next sections we will give a definition of the Bayesian Networks focusing on the simplest and computationally efficient of the Bayesian Networks versions, named \textit{\naive Bayes}. Finally, we will show our method for the challenge to evaluate the medium and the bid strategy.
\end{abstract}


%
\IEEEpeerreviewmaketitle

\copyrightstatement

\section{Bayesian Networks}
Bayesian Networks~(BN)~\cite{bayes} aid the reasoning and decision making in an uncertain environment. BN are built on top of a graphical model representing a set of random variables and their conditional independence. The learning phase in BN consists in retrieving the probabilities and the inference phase consists in computing the probabilities a posteriori by applying the Bayes rules. BN are powerful in terms of inference, they are able to produce a decision pattern with very little information. 
Very efficient algorithms exist for computing the probability a posteriori.\\
In the simplest BN model, all the variables are only dependent on the inferring variable. This model is called \naive Bayes (NB) and it is successfully applied to several domains, providing satisfactory results. 

\section{Solution}
We are going to focus our solution on \naive Bayes because it must be implemented in resource-constrained devices. In particular, we use the model to infer two variables: the condition of the network and the effectiveness of the bid. \\
We use both the variables to find the best strategy by applying a linear programming~\cite{comb} technique to maximize the gain and minimize the risk.

\subsection{Inferring the quality of the network}
Since the network is dynamic, we have to evaluate the stability of the path to where we would like to route the packet.
We use the information about the entire topology, provided and kept constantly updated by the OLSR~\cite{olsr} protocol, to train our \naive Bayes model and, consequently, evaluate the stability.\\

We define $\mathcal{N}$ as the set of nodes in the network. We assume $C_n$ is the event that our device is able to communicate with $n \in N$. Then, we compute $P(C_n)$ by recording, from the topology information, the number of times $n$ is reachable divided by the size of the sample.\\

The next step consists in computing the following variables:
\begin{itemize}
\item  Let $\mathcal{F}  \subseteq \mathcal{N}$ be the set of one-hop neighbors, $P(f | C_n)$ is the probability of using the neighbor $f \in \mathcal{F}$ for forwarding the packet,  the probability is computed by looking into the topology provided by OLSR.
\item Let $h \in \{1, .., |\mathcal{N}|\}$ be a random variable expressing the number of hops needed to reach the destination, $P(h | C_n)$  is the probability of sending the packet through a certain number of hops subjected to the fact that the packet successfully comes to the destination, this is used to evaluate the stability of the path depending on the number of necessary hops.
\end{itemize}

Inferring the quality of the network consist in inferring $P(C_n |Ê f, h)$ for all the nodes that offer the possibility to relay the message. We compute the inference by applying the Bayes rule.
\begin{equation}
\label{eq1}
P(C_n | f, h) = \frac{P(C_n)P(f | C_n)P(h | C_n) }{P(f)P(h)}
\end{equation}
In particular, we use Equation~\ref{eq1} to evaluate the probability of success in forwarding the message through all the neighbors devices. Then, we will use the results for the bid decision.

\subsection{Auctioneer Strategy}
We proceed in the same way as we have done in the previous subsection. We assume that we know the following information.
\begin{itemize}
\item $t \leq T$ is the timeout time (where $T$ is the maximum allowed timeout)
\item $b \leq B$ is the budget required from the bid (where $B$ is the maximum allowed budget)
\item $d \in \mathcal{N}$ is the final destination of the message.
\end{itemize}
We define the following variables:
\begin{itemize}
\item Let $s$ be the event of having a gain in the bid process, we compute the success by using the feedbacks provided by the Maniac Challenge 2013~\cite{manchal} framework.
\item Let $\lambda_1 \in \{0,1\}$ be the function that defines how much we can remove from $b$ in the offer we are proposing to the device who needs to forward the message.
\item Let $\lambda_2 \in \{0,1\}$ be the function that express our gain over $b$.
\end{itemize}
With the variables previously stated we compute the following probabilities:
\begin{itemize}
\item $P(t |Ês)$ is the probability of having a certain timeout subjected to the success.
\item $P(\lambda_1 |Ês)$ is the probability of remove a certain amount over $b$ subjected to having success in the bid.
\item $P(\lambda_2 |Ês)$ is the probability that we keep for us a certain amount of $b$ subjected to having success in the bid.
\item $P(d | s)$ is the probability of having a final destination $d$ subjected on having success.
\end{itemize}
Following the method mentioned in the Equation~\ref{eq1} we can compute the probability of success in the auction using Equation~\ref{eq2}.

\begin{equation}
\label{eq2}\
P(s | t, \lambda_1, \lambda_2, d) = \frac{P(s)P(t | s)P(d | s)P(\lambda_1 | s)P(\lambda_2 | s) }{P(t)P(d)P(\lambda_1)P(\lambda_2)}
\end{equation}

\subsection{Bidding Strategy}
The strategy is based on maximizing the gain and the probability of success, and minimizing the costs. So, we use the following linear programming expression (the value of $d$ is given in the information we receive from the bid requester).
 \begin{alignat*}{2}
    \text{maximize }     & P(C_d | f, h) +  \lambda_2 - \lambda_1 \\\\
    \text{subject to }  & P(s | t, \lambda_1, \lambda_2, d) \geq \theta_1&,\ & \lambda_1,\lambda_2 \in \{0,1 \}  \\
                    & P(C_d | f, h) \geq  \theta_2 &,\ & f \in N\\      
                & \lambda_1 + \lambda_2 \geq 0 
  \end{alignat*}
  Where $\theta_1$ and $\theta_2$ are the thresholds needed for evaluating the scenario.\\
 The linear programming expression previously shown could be easily solved with a local search procedure thanks to the small dimension of the search space.\\
In the case the linear programming expression does not return a feasible solution, the only action we can do consists in proposing a bid to forward the message via the wired backbone at the maximum price.

\section{Challenge Results and Analysis}


Compared to other strategies BManiac does not reach the desired results. 

The first issue we found stands in the time needed to the learning phase, BManic requires a priori set of sampling in order to retrieve the best inference and find a suitable solution. During the challenge the session times were too short to obtain enough sampling to achieve a good inference.

Moreover, we believe that another possible explanation BManic achieve poor results is in the fact that it tries to adapt his behavior depending on the offers coming from other devices instead to push them to modify their behavior.

We argue that BManiac could be easily adapted to different domains where cooperation is the centerpiece and it might reach good and fair results.

\section{Conclusion}
Our proposed solution for the Maniac Challenge 2013 is a heterogeneous algorithm based on both linear programming technique and Bayesian statistics. It is able to define a strategy for data offloading, which will provide good results in a cooperative environment with long lasting installations.





%

\end{document}